\documentclass[reprint,superscriptaddress,aps,prmaterials]{revtex4-2}

\usepackage[pdftex]{graphicx}
\usepackage{amsmath}

\begin{document}

\title{Bulk superconductivity in FeTe$_{1-x}$Se$_{x}$ via physicochemical pumping of excess iron}

\author{Lianyang Dong}
\affiliation{Materials Department, University of California, Santa Barbara, California 93106, USA}

\author{He Zhao}
\affiliation{Department of Physics, Boston College, Chestnut Hill, Massachusetts 02467, USA}

\author{Ilija Zeljkovic}
\affiliation{Department of Physics, Boston College, Chestnut Hill, Massachusetts 02467, USA}

\author{Stephen D. Wilson}
\email[Corresponding author: ]{stephendwilson@ucsb.edu}
\affiliation{Materials Department, University of California, Santa Barbara, California 93106, USA}

\author{John W. Harter}
\email[Corresponding author: ]{harter@ucsb.edu}
\affiliation{Materials Department, University of California, Santa Barbara, California 93106, USA}

\date{\today}

\begin{abstract}
The iron-based superconductor FeTe$_{1-x}$Se$_{x}$ has attracted considerable attention as a candidate topological superconductor owing to a unique combination of topological surface states and bulk high-temperature superconductivity. The superconducting properties of as-grown single crystals, however, are highly variable and synthesis dependent due to excess interstitial iron impurities incorporated during growth. Here we report a novel physicochemical process for pumping this interstitial iron out of the FeTe$_{1-x}$Se$_{x}$ matrix and achieving bulk superconductivity. Our method should have significant value for the synthesis of high-quality single crystals of FeTe$_{1-x}$Se$_{x}$ with large superconducting volume fractions.

\end{abstract}

\maketitle

\section{Introduction}

The iron-based superconductor FeTe$_{1-x}$Se$_{x}$ with chemical composition $x \approx 0.5$ possesses topological spin-helical Dirac surface states proximity-coupled to bulk $s$-wave superconductivity~\cite{wang2015,wu2016,xu2016}, realizing the celebrated Fu-Kane mechanism of topological superconductivity~\cite{fu2008}. Such a system hosts Majorana zero modes that obey non-Abelian exchange statistics, a crucial ingredient in topological quantum computing~\cite{nayak2008,alicea2012}. The relatively high superconducting critical temperature ($T_c = 14.5$~K) and facile growth of large single crystals make FeTe$_{1-x}$Se$_{x}$ a promising material platform to both generate and study Majorana fermions~\cite{zhang2018,wang2018,machida2019}. In as-grown crystals, however, excess iron may appear at interstitial sites in Fe$_{1+y}$Te$_{1-x}$Se$_{x}$. Depending on growth conditions and the starting Te:Se ratio, the excess iron concentration $y$ can reach values exceeding 10\%. While a starting Te:Se ratio of 1:1 has been shown to yield bulk superconducting crystals with $y\approx 0$, moving away from this ratio is known to rapidly degrade superconductivity via the promotion of interstitial iron impurities~\cite{sales2009,yeh2009,bendele2010,dong2011,cao2011,sun2014}. Furthermore, a gradated iron composition may also develop within a single crystal boule, rendering the superconductivity inhomogeneous across the crystal.

To remedy this problem, a number of single-step post-growth annealing solutions have been proposed to reduce $y$. These methods have been performed in both an oxidizing environment and in an overpressure of Te/Se vapor~\cite{mizuguchi2010,kawasaki2012,sun2013a,sun2013b,zhou2014,xu2018,sun2019}. Nevertheless, their application remains limited due to a strong dependence on crystal morphology as well as uncertainties concerning the presence of superconductivity within the surface-shielded core of bulk crystals. The mechanisms underlying these annealing methods and their ability to tune iron content throughout the bulk of Fe$_{1+y}$Te$_{1-x}$Se$_{x}$ crystals remain open questions. Of particular intrigue is the use of oxygen containing environments to anneal Fe$_{1+y}$Te$_{1-x}$Se$_{x}$ crystals. The propensity of O$_2$ to oxidize iron provides a strong chemical driving force for removing interstitial iron from the FeTe$_{1-x}$Se$_{x}$ matrix, but the extent to which this process is diffusion-limited prior to the process decomposing the FeTe$_{1-x}$Se$_{x}$ lattice itself is unclear. In particular, the limitations on how much interstitial iron can be cleanly removed from the matrix and over what length scale (i.e. crystal thickness) have yet to be determined.

In this paper, we explore the use of oxygen as an annealing agent to remove excess iron from Fe$_{1+y}$Te$_{1-x}$Se$_{x}$ crystals. For crystals greater than $\sim 30$~$\mu$m in thickness, we find that a single annealing step is insufficient to produce bulk superconductivity, instead inducing superconductivity only on the outer shell of the sample exposed to the oxidizing environment. To mitigate this limitation, we develop a cyclical post-annealing method to gradually improve the superconducting volume fraction by physicochemically pumping excess iron out of the material, eventually achieving fully bulk superconductivity. Specifically, by using repeated cycles of oxygen annealing, acid etching, and vacuum annealing, we demonstrate that $y\approx 0.08$ can be reasonably removed from the bulk of an Fe$_{1+y}$Te$_{1-x}$Se$_{x}$ single crystal with thickness approaching several hundred microns. We envision this technique will be useful for the tailored experimental exploration of the candidate topological superconducting state in this material system.

\begin{figure*}[t]
\includegraphics{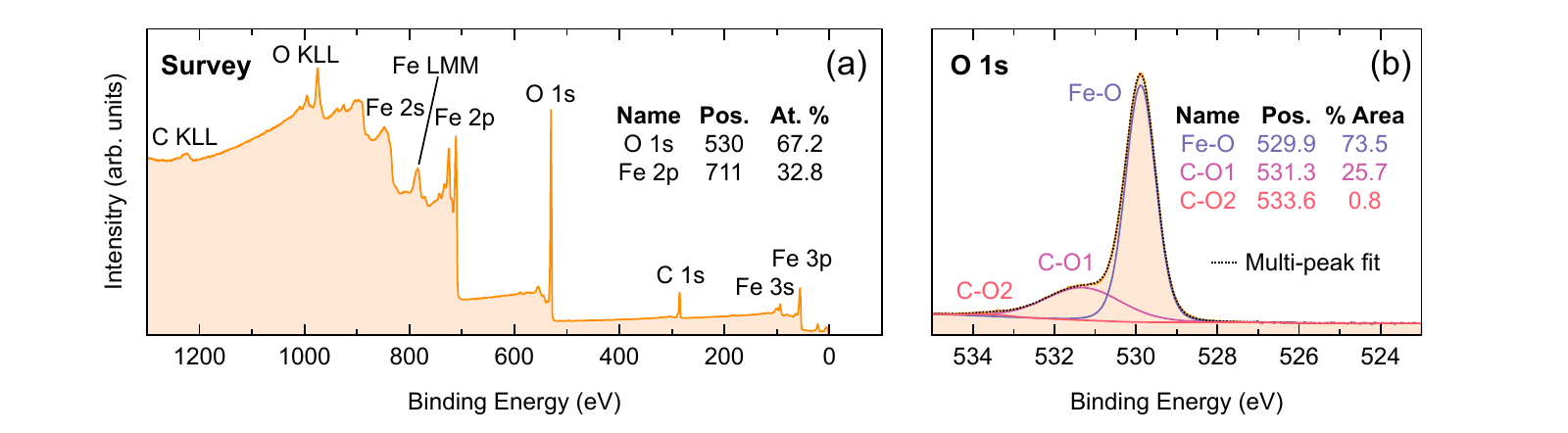}
\caption{\label{FigXPS} XPS measurements of an air-annealed sample of Fe$_{1+y}$Te$_{1-x}$Se$_{x}$. (a)~Survey spectrum showing the presence of iron, oxygen, and carbon on the surface. Analysis of O~1$s$ and Fe~2$p$ peaks gives an initial Fe:O atomic ratio of 1:2.05. (b)~High-resolution scan of the O~1$s$ peak, from which it is determined that only 73.5\% of oxygen is bonded to iron, with the remainder coming from various carbon compounds on the surface. This yields a corrected Fe:O ratio of 1:1.51 (Fe$_2$O$_3$).}
\end{figure*}

\section{Experimental Details}

Single crystals with nominal composition FeTe$_{0.55}$Se$_{0.45}$ were grown using a self-flux method similar to previous reports~\cite{okazaki2012,sun2013b}. Starting materials Fe~(99.99\%), Te~(99.999\%), and Se~(99.999\%) were well mixed in the molar ratio 1:0.55:0.45 and double-sealed in quartz ampoules evacuated to high vacuum ($\sim 10^{-5}$~mbar). The sealed ampoules were placed in a furnace and heated slowly to 1100~${^\circ}$C, held for 20~hours, cooled to 650~${^\circ}$C at a rate of 2~${^\circ}$C per hour, and then cooled to 400~${^\circ}$C at a rate of 5~${^\circ}$C per hour. The furnace was then powered off and cooled down to room temperature. Single crystals were removed from the solidified flux by breaking the quartz ampoules in air. 

X-ray fluorescence (XRF) measurements were performed using a Rigaku ZSX Primus IV (WDXRF). A 1~mm probe diameter was chosen for analysis of thin plate single crystals with sizes up to 6~mm. X-ray photoelectron spectroscopy (XPS) measurements were conducted using a Thermo Fisher Escalab Xi+ instrument. A monochromatic Al~K$\alpha$ x-ray source was used for all samples. Pass energies were 100~eV for survey scans and 20~eV for high-resolution scans. Samples were measured without cleaving in vacuum in order to study the thin blue oxide layer on the sample surface, resulting in the detection of some surface carbon compounds which will be discussed later. Magnetization measurements were performed using a Quantum Design MPMS3 SQUID magnetometer. Crystals were mounted on a quartz paddle with GE varnish and magnetization data was collected using a 10~Oe field. Powder samples (crushed crystals) were placed inside of a plastic capsule mounted onto a brass half-tube sample holder.

Scanning tunneling microscopy (STM) measurements were performed on single crystals cleaved under ultra-high vacuum at 80~K and immediately inserted into the STM head in order to expose a pristine surface free of contaminants. STM data was acquired at a base temperature of 4.5~K using a commercial Unisoku USM1300 system. Spectroscopic measurements were taken using standard lock-in techniques at 915~Hz. STM tips were home-made chemically etched tungsten. Tip sharpness was first evaluated under an optical microscope before the tip was annealed in vacuum to a bright orange color and inserted into the STM. 

\section{XRF and XPS Analysis}

As-grown crystals of Fe$_{1+y}$Te$_{1-x}$Se$_{x}$ were initially analyzed by XRF. Elemental analysis of multiple samples consistently showed a tellurium-rich and iron-rich composition, with an average Fe:Te:Se ratio of 1.08:0.59:0.41. Selenium is more volatile than tellurium, and it is likely that a small fraction evaporated from the melt and deposited onto the walls of the quartz ampoule during growth. Notably, the Te:Se ratio found in these crystals rests on the boundary where excess iron begins to be incorporated during growth and bulk superconductivity vanishes~\cite{sales2009}. This is the regime where post-growth treatment of crystals becomes necessary. This as-grown Fe$_{1.08}$Te$_{0.59}$Se$_{0.41}$ composition demonstrates the need for our approach to inducing bulk superconductivity.

To explore the effects of annealing in an oxidizing environment, a small crystal was cleaved from a larger single crystal boule of Fe$_{1.08}$Te$_{0.59}$Se$_{0.41}$. Crystals cleave with shiny metallic surfaces, but the color of the crystal surface is known to turn blue after post-annealing in air or pure O$_2$~\cite{dong2011,sun2014}. This suggests that the excess iron removed from the bulk crystal during the annealing step remains on the crystal surface as an oxide. To verify this, we carried out XPS measurements on the blue surface layer of an air-annealed sample (300~$^\circ$C for 4~hours). As shown in Fig.~\ref{FigXPS}(a), the elements iron, oxygen, and carbon were detected, indicating that the blue surface consists of iron oxide and some carbon compounds. The overall Fe:O ratio was measured to be 1:2.05, but the spectral shape of the O~$1s$ peak shown in Fig.~\ref{FigXPS}(b) indicates that approximately 26.5\% of the oxygen was bonded to carbon, with the remainder bonded to iron, giving a corrected Fe:O ratio of approximately 1:1.51. This analysis shows that the surface layer is mainly Fe$_2$O$_3$, with the blue hue coming from thin film optical interference effects between the crystal surface and the outer oxide layer. The fact that the color is always blue suggests that the oxide layer that forms under moderate oxidizing conditions is self-limiting. As we will discuss later, the Fe$_2$O$_3$ surface layer that develops can be removed by briefly etching in a solution of hydrochloric acid without damaging the underlying Fe$_{1+y}$Te$_{0.59}$Se$_{0.41}$ crystal.

\section{Single Post-Annealing Step}

\begin{figure}[t]
\includegraphics{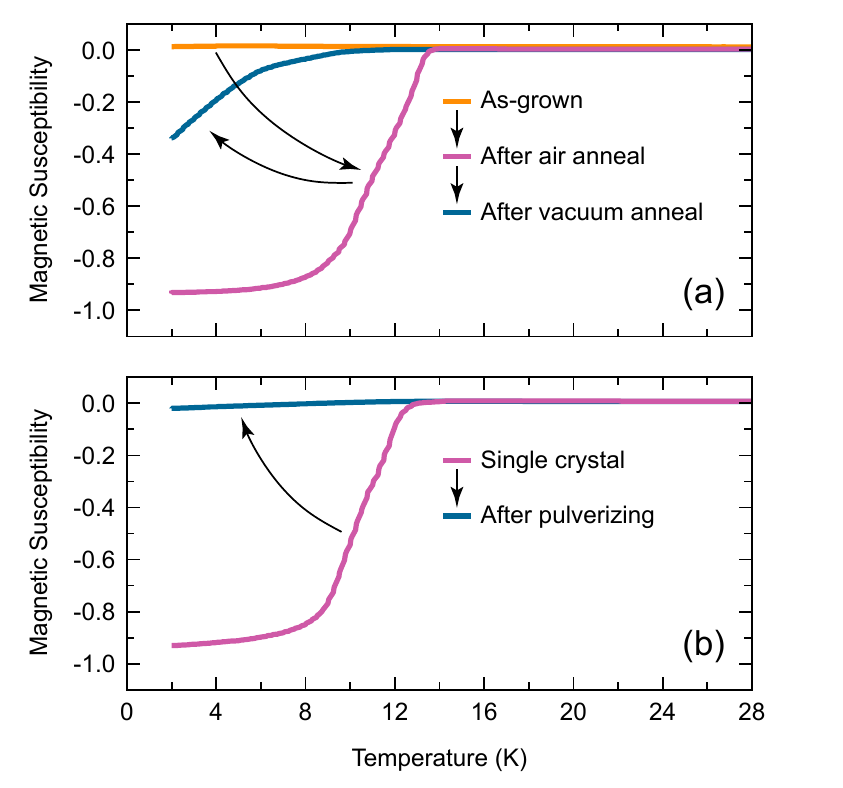}
\caption{\label{FigSingleCycle} Magnetic susceptibility measured during a single post-annealing cycle for samples with thickness $\sim 75$~$\mu$m. (a)~As-grown samples do not exhibit signs of superconductivity, but after annealing in air, a large diamagnetic signal is measured. The diamagnetism is suppressed by subsequent vacuum annealing. (b)~The apparent superconducting volume fraction of an air-annealed sample is significantly reduced after pulverizing.}
\end{figure}

As shown in Fig.~\ref{FigSingleCycle}(a), the magnetic susceptibility of an as-grown Fe$_{1.08}$Te$_{0.59}$Se$_{0.41}$ crystal of thickness 75~$\mu$m does not exhibit a superconducting diamagnetic transition because the excess iron completely suppresses superconductivity. After annealing in air at 300~$^\circ$C for 4~hours, a clear superconducting transition is observed at $T_c \approx 14$~K with an apparent large superconducting volume fraction. This large shielding fraction is due to a shell effect as shallow excess iron is pulled from the crystal and oxidized only at the surface. If this same crystal is then sealed under vacuum and annealed at 400~$^\circ$C for 24~hours, the apparent superconducting volume fraction decreases significantly. This is due to the vacuum anneal causing the remaining excess iron in the bulk of the crystal to diffuse to the surface. To confirm this hypothesis, an air-annealed sample with a clear superconducting transition was pulverized into a powder, after which the superconducting volume fraction dropped to almost zero, as shown in Fig.~\ref{FigSingleCycle}(b). Both observations strongly suggest that the superconductivity that appears after annealing in air occurs at the surface only and that the apparent superconducting volume fraction determined by the Meissner effect is not indicative of the bulk.

\section{Physicochemical Pumping}

\begin{figure}[t]
\includegraphics{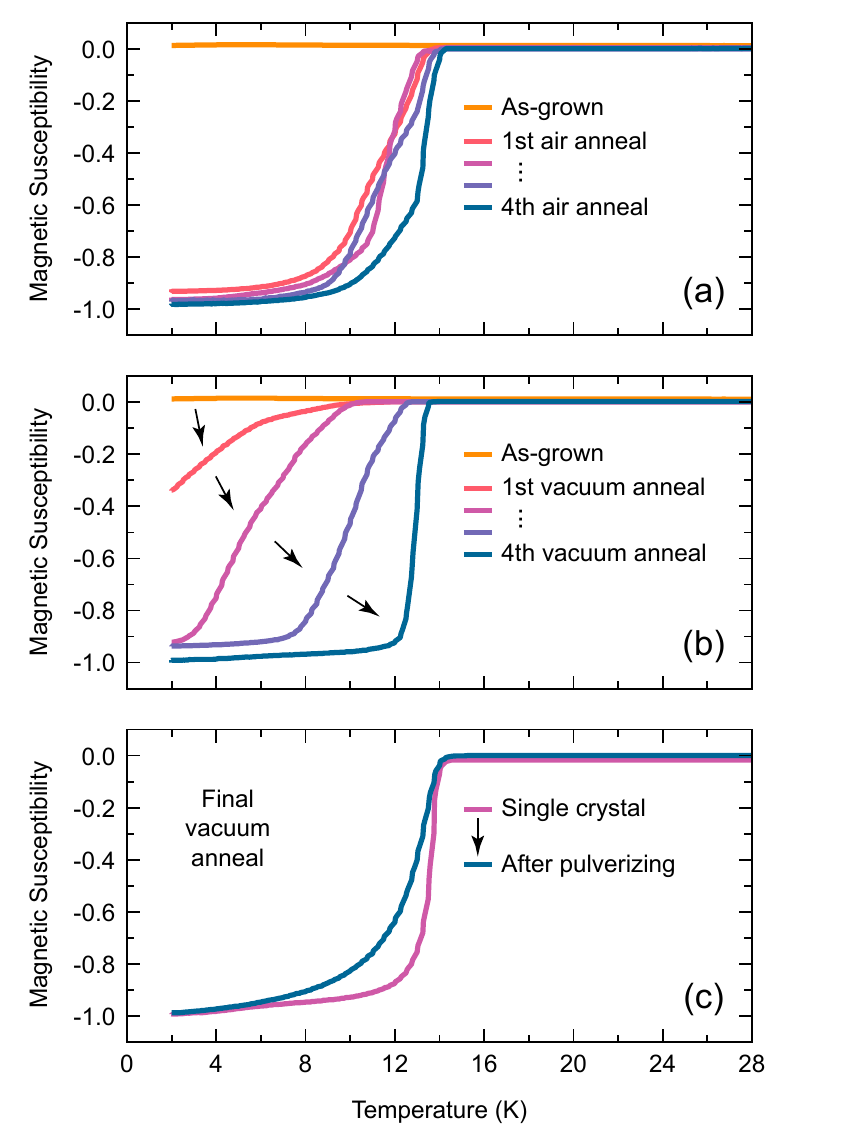}
\caption{\label{FigManyCycles} Evolution of magnetic susceptibility after multiple annealing cycles for samples with thickness $\sim 75$~$\mu$m. (a)~Annealing in air consistently yields apparent superconducting volume fractions above 90\%. (b)~A steady cycle-to-cycle improvement in volume fraction is observed after the vacuum annealing steps. (c)~After the final vacuum annealing step, pulverization of a single crystal no longer reduces the apparent superconducting volume fraction, indicating bulk superconductivity has been achieved.}
\end{figure}

\begin{figure}[t]
\includegraphics{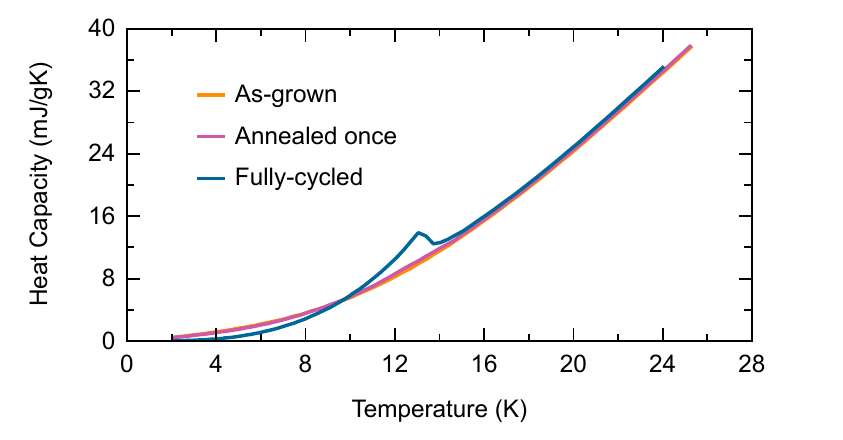}
\caption{\label{FigHeatCapacity} Heat capacity versus temperature for an as-grown (non-superconducting) crystal, a crystal that has undergone a single post-annealing step and exhibits a large diamagnetic transition, and a fully-cycled crystal. Annealing once results in a small change in heat capacity at the critical temperature, likely resulting from surface superconductivity, but only fully-cycled crystals show a large heat capacity anomaly indicative of bulk superconductivity.}
\end{figure}

Here we show that to obtain fully bulk superconductivity, multiple annealing cycles are required. Each cycle consists of three steps: (1) annealing in air to oxidize shallow excess iron impurities, (2) etching in hydrochloric acid to remove the resulting iron oxide, and (3) annealing in vacuum to redistribute or ``pump'' iron impurities from the bulk to the surface via diffusion. Carrying out several cycles of this procedure on the same sample, we observed that an apparent superconducting volume fraction above 90\% is consistently observed after each anneal in air, as shown in Fig.~\ref{FigManyCycles}(a). On the other hand, after subsequent annealing in vacuum the superconducting volume fraction steadily increased from cycle to cycle to nearly 100\%, as shown in Fig.~\ref{FigManyCycles}(b). This behavior is expected if a small amount of excess iron is removed from the bulk after the completion of each cycle, thereby gradually improving the superconducting properties until a sharp transition and bulk superconductivity are reached. To verify this end point, after performing several annealing cycles, the magnetic susceptibilities of a sample measured before and after pulverization were compared, as shown in Fig.~\ref{FigManyCycles}(c). The similar near-100\% superconducting volume fractions of the powder and single crystal confirm that bulk superconductivity is achieved after several annealing cycles. Further verification is obtained through bulk-sensitive heat capacity measurements, where a large anomaly at $T_c$ is only observed after several cycles, as shown in Fig.~\ref{FigHeatCapacity}.

\begin{figure}[t]
\includegraphics{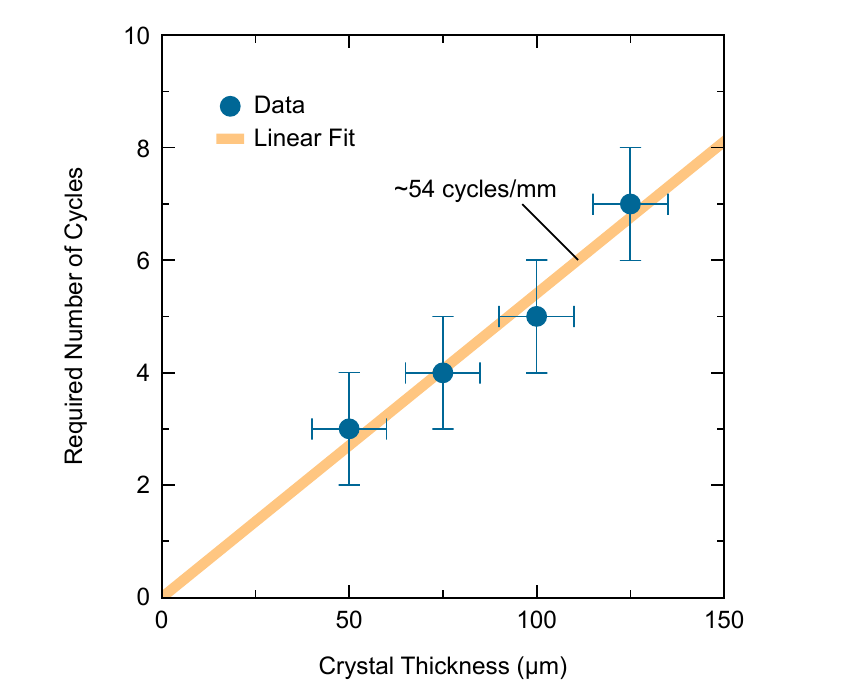}
\caption{\label{FigThicknessDependence} Number of post-annealing cycles required to reach bulk superconductivity versus crystal thickness. An approximately linear dependence is observed with a proportionality constant of $\sim 54$~cycles per millimeter of thickness.}
\end{figure}

In Fig.~\ref{FigThicknessDependence} we plot the approximate number of cycles required to reach bulk superconductivity (no change in volume fraction after vacuum annealing) versus crystal thickness. With increasing thickness, the number of required cycles increases, following a linear trend. This functional relationship supports the hypothesis that a fixed amount of excess iron per unit area is pumped from the sample during each cycle, as would be expected for self-limiting surface oxidation. In particular, a thicker crystal requires the removal of a proportionally larger amount of iron before bulk superconductivity is achieved, and therefore requires a proportionally larger number of post-annealing cycles. This means that the cyclical post-annealing technique is rate-limited by the oxidation step. The proportionality constant extracted from the data suggests $\sim 54$~post-annealing cycles are required for each millimeter of crystal thickness. XRF measurements of fully-cycled crystals exhibiting bulk superconductivity showed that the iron content had decreased to $(1+y)\approx 0.98-0.99$ following the final annealing step. This demonstrates that interstitial iron levels as high as $y = 0.08$ can be successfully removed from several hundred $\mu$m thick crystals in a manageable number ($< 10$) of cycles. For example, Table~\ref{TableXRF} shows a steady decrease in iron content for a sample that underwent four annealing cycles.

\begin{table}[h]
\begin{tabular}{c|c}
Step & Fe ($1+y$) \\
\hline
As-grown & 1.063 \\
Cycle 1 & 1.040 \\
Cycle 2 & 1.030 \\
Cycle 3 & 1.013 \\
Cycle 4 & 0.983
\end{tabular}
\caption{\label{TableXRF} Iron content measured by XRF for a sample undergoing several physicochemical pumping cycles. Each value is the average of four randomly-chosen spots on the sample surface.}
\end{table}

\begin{figure}[t]
\includegraphics{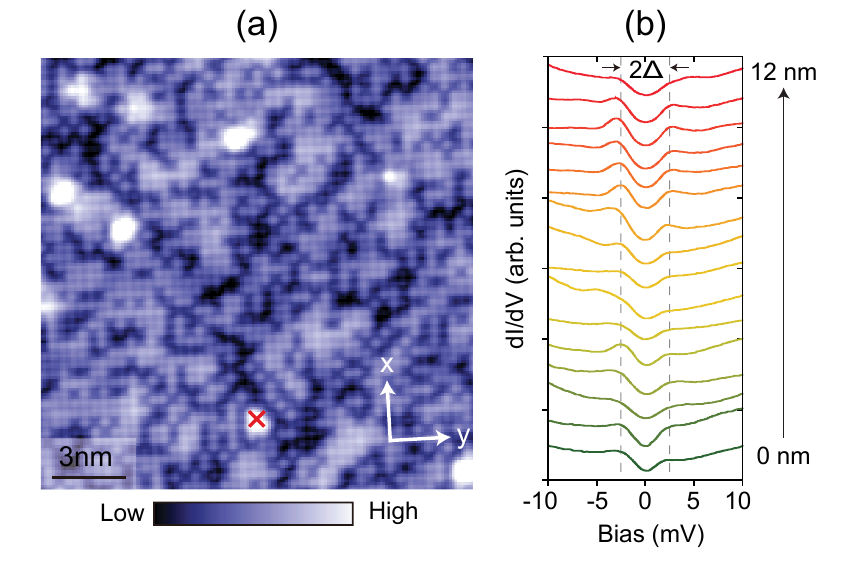}
\caption{\label{FigSTM} STM data from a cleaved surface of FeTe$_{0.59}$Se$_{0.41}$ measured at 4.5~K. (a)~Representative topographic image showing tellurium atoms (bright spots) and selenium atoms (dark spots) randomly distributed on a square lattice. Counting the spots yields a Te:Se ratio of 0.61:0.39. Also apparent are brighter spots representing interstitial iron impurities. (b)~$dI/dV$ spectra taken along a cut extending away from the interstitial iron impurity marked by a red cross in panel (a), showing a superconducting energy gap of $\sim 2.5$~meV. STM setup conditions: $I_\mathrm{set} = 100$~pA; $V_\mathrm{sample} = -10$~mV; $V_\mathrm{exc} = 0.2$~mV (zero-to-peak).}
\end{figure}

To further explore the local homogeneity and bulk superconductivity of a fully-cycled crystal, we characterized a bulk-superconducting sample using STM. An STM topograph is shown in Fig.~\ref{FigSTM}(a), depicting a chalcogen-atom-terminated cleaved surface with randomly distributed bright and dark spots on a square lattice, corresponding to tellurium and selenium atoms, respectively. Counting the spots gives a Te:Se ratio of 0.61:0.39, consistent with the XRF results. The topograph also shows several brighter spots scattered on the surface, corresponding to interstitial iron impurities centered at the midpoint of four neighboring Te/Se atoms. This excess iron level is very close to the nominal $y = 0$ value, with only 5 interstitial iron sites identified among more than 2000 unit cells. Shown in Fig.~\ref{FigSTM}(b) are $dI/dV$ tunneling spectra. All spectra exhibit a large superconducting energy gap of $\sim 2.5$~meV, demonstrating that the superconducting state is uniform across the sample. We note here that the STM data do not resolve the smaller gap (1.5 -- 2~meV) or zero bias conductance peak typically seen in other STM data~\cite{yin2015}. These observations could be due to the different Te:Se ratio, where at higher ratios the sample may no longer be topological.

\begin{figure*}[t]
\includegraphics{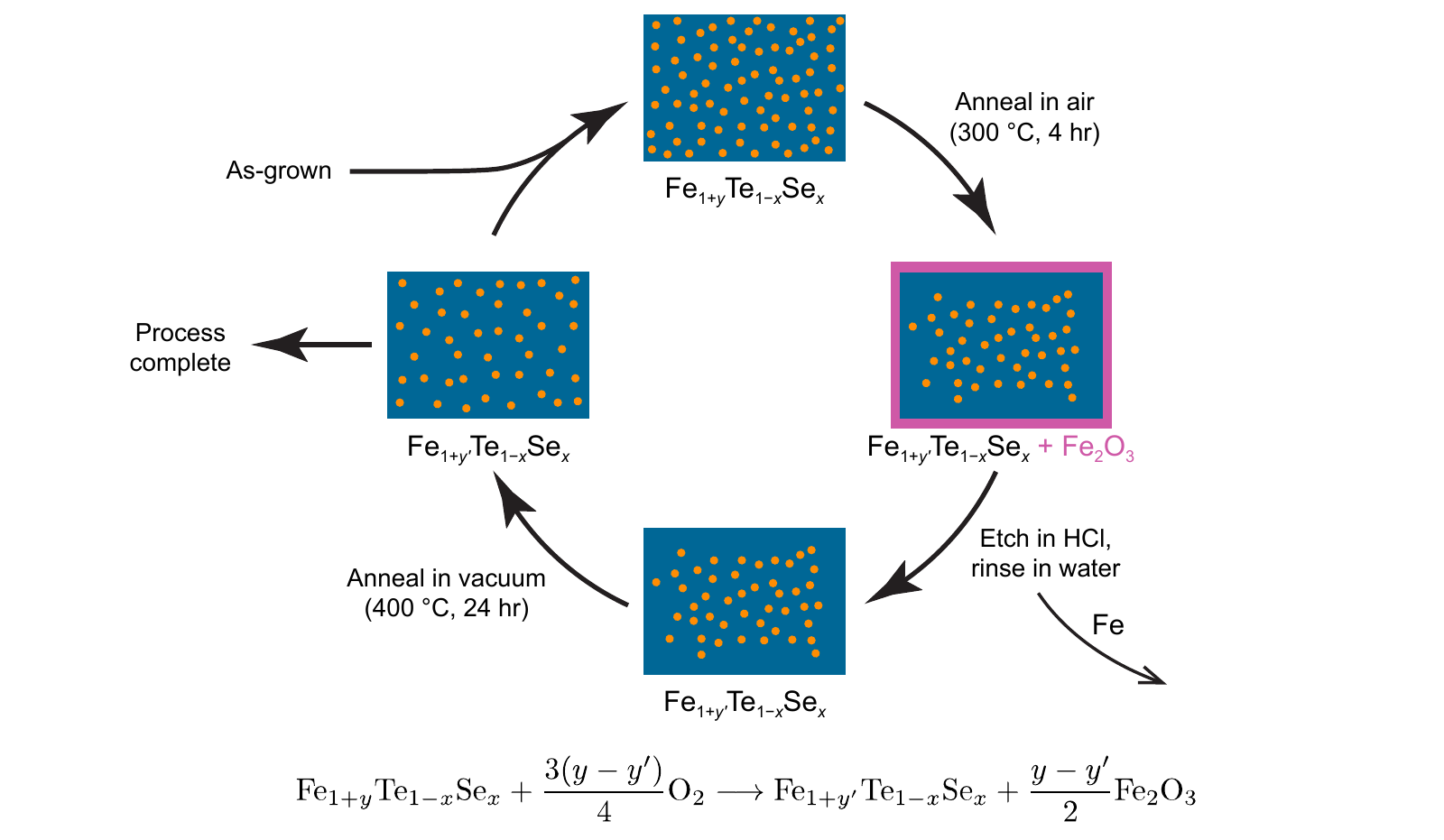}
\caption{\label{FigProcess} Illustration of the cyclical physicochemical pumping procedure. Orange dots represent excess iron impurities. Due to the self-limiting nature of the oxidation step, after each cycle a fixed amount of excess iron is removed from the crystal. The process may be considered complete when the vacuum annealing step no longer changes the volume fraction determined from magnetic susceptibility measurements.}
\end{figure*}

\section{Discussion}

A number of differing approaches have been proposed for tuning the excess iron content within Fe$_{1+y}$Te$_{1-x}$Se$_{x}$, ranging from vacuum annealing~\cite{dong2011}, to oxygen annealing~\cite{sun2013a,sun2013b}, to annealing in tellurium vapor~\cite{xu2018}. Depending on the specific processing technique used, these approaches may suffer from diffusion-limited removal of iron deep inside bulk crystals, sample degradation under heat treatment, or an inhomogeneous distribution of iron across the sample. The iterative gradual removal of interstitial iron presented here mitigates these challenges. Our cyclical post-annealing process involves iron being physicochemically pumped from the bulk to the surface and then removed via a three-step process of oxidation of shallow iron to Fe$_2$O$_3$, acid etching to physically remove the surface oxide, and thermally-activated diffusion to redistribute the remaining iron in the crystal. The process is illustrated in Fig.~\ref{FigProcess}. The real superconducting volume fraction is gradually improved after each annealing cycle, and bulk superconductivity can typically be achieved after several cycles.

While we have chosen reasonably gentle oxidizing conditions for the first step of the cycle (300~$^\circ$C in air), this step may be further optimized to reduce the required total number of cycles for a given sample thickness. Earlier single-step annealing studies of thin Fe$_{1+y}$Te$_{1-x}$Se$_{x}$ crystals suggest this to be the case~\cite{mizuguchi2010,sun2014}. Intriguingly, tailoring the multistep process to realize finite $y$ values within bulk single crystals may also allow for new avenues to study both the competing magnetic states (stabilized via interstitial iron~\cite{yin2015,viennois2010}) and superconductivity as well as the impact of finite $y$ on topological superconductivity. Exploring this process for tailored $y$ values and with varying Te:Se ratios are appealing next steps.

\section{Conclusion}

FeTe$_{1-x}$Se$_{x}$ is a promising material platform for studying topological superconductivity and Majorana fermions. In as-grown crystals, however, superconductivity is usually suppressed by excess iron impurities. While these impurities may be removed by annealing in an oxidizing environment, we demonstrate that for crystals with thicknesses greater than $\sim 30$~$\mu$m, a single annealing step is insufficient to remove iron from the bulk. To address this, we present a cyclic multistep annealing/etching process for the production of bulk superconducting crystals with $y \approx 0$. Our method should have significant value for the synthesis of high-quality single crystals of FeTe$_{1-x}$Se$_{x}$ with large superconducting volume fractions at Te:Se ratios away from the commonly studied 1:1 values. We speculate that the general physicochemical pumping procedure developed here may also be applicable to the growth of other material systems where as-grown impurities must be removed in order to improve physical properties.

\section*{Acknowledgments}

This work was supported by the Materials Research Science and Engineering Centers (MRSEC) program of the National Science Foundation through Grant No.~DMR-1720256 (Seed Program). I.Z. acknowledges support from the Army Research Office through Grant No.~W911NF-17-1-0399.

\end{document}